\documentclass[12pt]{article}
\usepackage{epsfig}\parskip 5pt plus 1pt
\usepackage{amssymb}
\usepackage{amsmath}
\newcommand{\be}{\begin{equation}}
\newcommand{\ee}{\end{equation}}
\newcommand{\bea}{\begin{eqnarray}}
\newcommand{\eea}{\end{eqnarray}}

\def\simge{\mathrel{%
   \rlap{\raise 0.511ex \hbox{$>$}}{\lower 0.511ex \hbox{$\sim$}}}}
\def\simle{\mathrel{
   \rlap{\raise 0.511ex \hbox{$<$}}{\lower 0.511ex \hbox{$\sim$}}}}

\begin{document}
\thispagestyle{empty}
\vspace*{1cm}
\begin{center}
{\Large{\bf Particle acceleration in sub-cycle optical cells} }\\

\vspace{.5cm}
F.~Terranova$^{\rm a}$ \\
\vspace*{1cm}
$^{\rm a}$ Dep. of Physics, Univ. of Milano-Bicocca and INFN, 
Sezione di Milano-Bicocca, Milano, Italy \\
\end{center}

\vspace{.3cm}
\begin{abstract}
\noindent
A single laser pulse with spot size smaller than half its wavelength
($w_0 < \lambda/2$) can provide a net energy gain to ultra-relativistic
particles. In this paper, we discuss the properties of an optical cell
consisting of $N$ sub-cycle pulses that propagate in the direction
perpendicular to the electron motion. We show that the energy gain
produced by the cell is proportional to $N$ and it is sizable even for
$\mathcal{O}(1\mathrm{~TW})$ pulses. 
\end{abstract}

\vspace*{\stretch{2}}
\begin{flushleft}
\end{flushleft}

\newpage
 
Modern conventional accelerators transfer only a small amount of
energy to particles per unit cell.  Large energy gradients are
achieved combining the cells in periodic structures and phase locking
the particle motion to the RF cavities. On the other hand, phase
slippage is difficult to be avoided in laser-based accelerators and
acceleration is achieved by a few beam-pulse interactions but, in this
case, the energy transfer per pulse can be very large due to the huge
electric fields available in lasers. Unlike RF-based accelerators, the
interaction time $\tau$ of the particle with the laser is much longer
that the period $T$ of the laser fields. Such a long interaction time
causes several drawbacks.  Plasma-based
accelerators~\cite{esarey_review} tend to be prone to instabilities
when $\tau \gg T$. This difficulty has been overcome by the
development of high-power sub-ps lasers and resulted into large energy
gains and beams of unprecedented
quality~\cite{Banerjee:2013kwa}. Similarly, laser accelerators in
vacuum~\cite{acc_vacuum,esarey_1995} require special geometrical
configurations of the laser pulse to bypass the limitations imposed by
the Lawson-Woodward theorem~\cite{lawson_woodward} and reach a finite
energy gain even if $\tau \gg T$~\cite{Shao:2011rp}. These challenges
have not been overcome yet, and vacuum-based techniques are much less
developed than laser-plasma accelerators. As discussed in this paper,
however, the impressive progresses in sub-cycle laser technology
achieved in the last decade could boost vacuum acceleration in a way
similar to what sub-ps lasers have done in the past for plasma-based
accelerators.

\begin{figure}
\centering\includegraphics[width=0.7\textwidth]{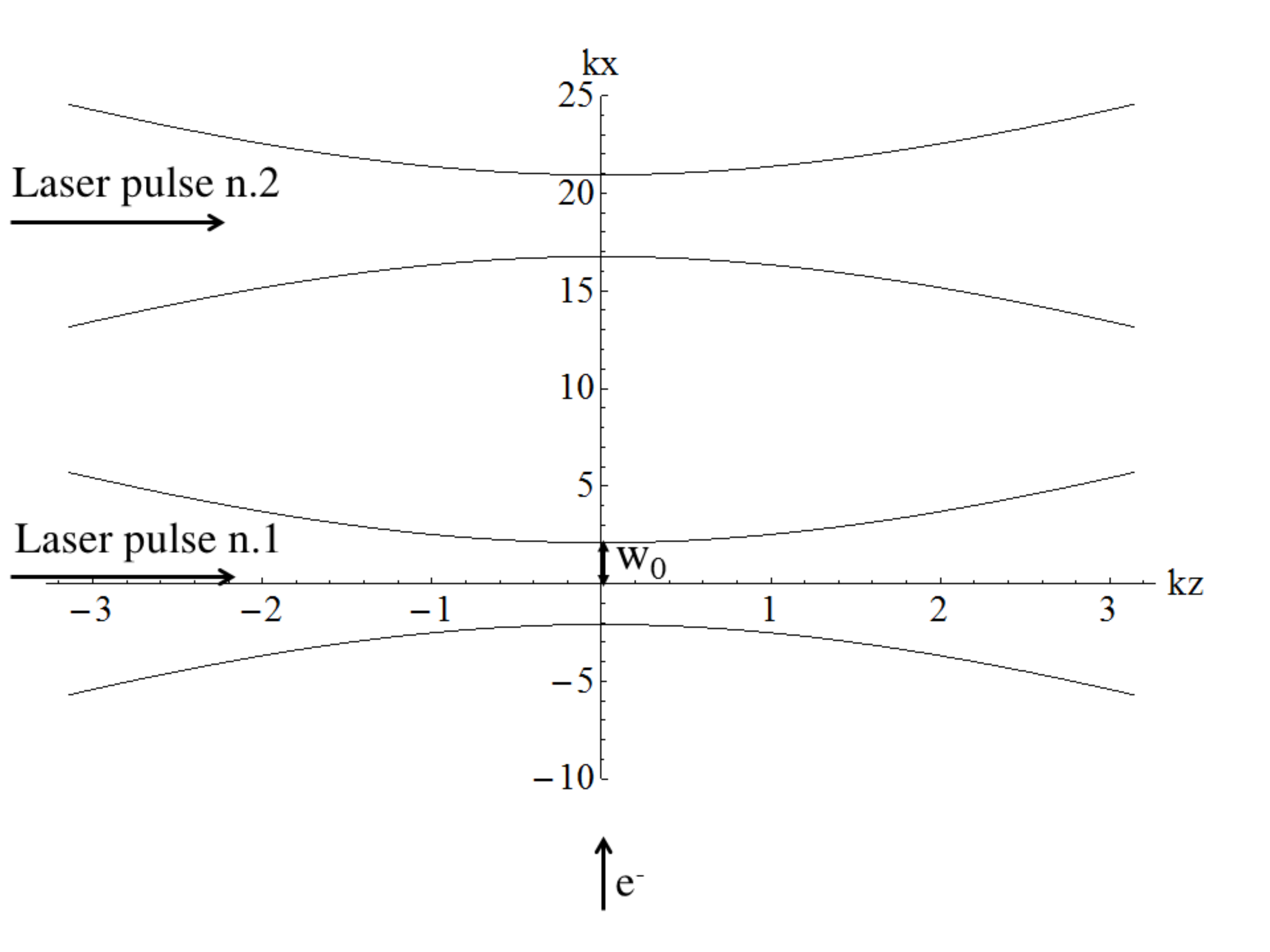} 
\caption{A $N$=2 optical cell. The laser pulses propagate along the $z$
  axis and the particles are injected along $x$. Distances are
  expressed in units of $\lambda$: $kz$ (or $kx$) $= 2\pi$ corresponds
  to a shift of $\lambda$ in the $z$ ($x$) direction. $w_0$ is the
  beam spot (see text). }
\label{fig:scheme}
\end{figure}

Attosecond optics allows for the focusing of pulses to spot sizes
significantly smaller than $\lambda$~\cite{dorn_2003}, and for
creation~\cite{goulielmakis_science_2008,krauss_naturephotonics_2009},
manipulation and regeneration of isolated sub-cycle
pulses~\cite{chang}. In addition, accelerators based on sub-cycle
pulses can be operated in the $\tau < T$ regime: particles that are
injected nearly perpendicularly with respect to the pulse propagation
axis (see Fig.~\ref{fig:scheme}) will experience only a fraction of
the oscillating electric field and, if properly phased, they will
acquire a net energy gain when they cross the focus of the pulse. A
set of $N$ sub-cycle pulses can be employed to transfer a fixed amount
of energy to the particles through $N$ beam-pulse interactions. If the
overall energy gain is constant and the beam is stable in phase-space,
then the $N$-pulse optical lattice acts as a single accelerating unit
(optical cell) with a well defined transfer function. Particles
crossing an optical cell will experience an energy gain that, in
general, will depend on the initial position of the particle;
moreover, the particles will be deflected in the transverse plane
($y-z$ in Fig.~\ref{fig:scheme}) by the longitudinal electric field
$E_z$ and by the ponderomotive force. \
In this paper we study
the accelerating properties of the optical cell and show that, in
suitable conditions, this device can provide an energy gain
proportional to $N$ and is described by a transfer function similar to
the one of a defocusing lens.

The advantage of an optical cell compared to standard RF cells resides
mostly in the intensity of the electric field, which can exceed the RF
field by more than six orders of magnitude. In addition, the size of
the cell in the $x$ direction is ${\cal O}(N\lambda)$ and, therefore,
is suitable for tabletop acceleration. On the other hand, the
acceptance of the optical cell, i.e. the region in the
transverse plane where particles can experience acceleration, is small
(${\cal O}(\lambda) \sim 1-10$~$\mu$m) and comparable to the vertical
beam size of high energy accelerators or free-electron
lasers~\cite{PDG} for optical or IR lasers.

In this paper, we consider an optical cell in its simplest
configuration (Fig.~\ref{fig:scheme}): a set of $N$ pulses reflected
back and forth along the $z$ axis and focused to an area $< \lambda^2$
at $z=z_0$ ($z_0=0$ in Fig.~\ref{fig:scheme}). The length of the pulse
along $z$ is assumed to be $\gg \lambda$ so that the light beam has a
stationary focus at $z=z_0$. The pulse profile at $z$ is circular with
radius $w(z) = w_0 \sqrt{1+(\frac{z-z_0}{z_R})^2}$, $z_R=k w_0^2/2$
being the Rayleigh length of the pulse and $k=2 \pi/\lambda$ the
wavenumber of the light at central frequency. The e.m. fields in the
proximity of the focus can be derived following the approach of Davis
and McDonald~\cite{davis,mcdonald}. The light beam is modeled
employing a vector potential $\mathbf{A}$ linearly polarized along
$x$ and expressing $\mathbf{A}$ as a perturbative expansion of the
diffraction angle $\epsilon = w_0/z_R$. Such treatment, which is
consistent as long as $\epsilon<1$ ($w_0>\lambda/\pi$), provides the
electric and magnetic fields in the proximity of the focus as a
function of the transformed variables:
\be 
\xi = x/x_0 \ , \hspace{0.5cm} v = y/w_0 \ , \hspace{0.5cm} \zeta
= z/z_R \ , \hspace{0.5cm} r^2 = x^2+y^2 \ , \hspace{0.5cm} \rho = r/w_0
\ee
and of the plane wave phase $\eta = \omega t - kz$. 
At next-to-leading order~\cite{Salamin:2002dd} the fields are:
\be
E_x = E_0 \frac{w_0}{w} e^{ -\frac{r^2}{w^2} } \left\{ S_0
  + \epsilon^2 \left[ \xi^2 S_2 - \frac{\rho^4 S_3}{4} \right] + 
\mathcal{O}(\epsilon^4) \right\}  
\label{eq:ex}
\ee
\be
E_y =  E_0 \frac{w_0}{w} e^{ -\frac{r^2}{w^2} } \xi v 
 \left\{ \epsilon^2 S_2 + \mathcal{O}(\epsilon^4) \right\} 
\ee
\be
E_z =  E_0 \frac{w_0}{w} e^{ -\frac{r^2}{w^2} } \xi \left\{
\epsilon C_1 + \epsilon^3 \left[ -\frac{C_2}{2} +\rho^2 C_3 - 
\frac{\rho^4 C_4}{4} \right] + \mathcal{O}(\epsilon^5)  \right\}
\ee
 \be
B_x = 0
\ee
\be
B_y =  E_0 \frac{w_0}{w} e^{ -\frac{r^2}{w^2} }  
\left\{ \epsilon^2 \left[ \frac{\rho^2 S_2}{2}
- \frac{\rho^4 S_3}{4} \right] 
+ \mathcal{O}(\epsilon^4)  \right\} 
\ee
\be
B_z = E_0 \frac{w_0}{w} e^{ -\frac{r^2}{w^2} } v \left\{
  \epsilon C_1 + \epsilon^3 \left[ \frac{C_2}{2} + \frac{\rho^2
      C_3}{2} - \frac{\rho^4 C_4}{4} \right] + \mathcal{O}(\epsilon^4)  
\right\}
\label{eq:bz}
\ee
where
\begin{gather}
S_n = \left( \frac{w_0}{w} \right)^n sin(\psi+n \psi_G) \hspace{1cm}
C_n = \left( \frac{w_0}{w} \right)^n cos(\psi+n \psi_G) \ .
\label{eq:sin}
\end{gather}
The phases entering Eq.~\ref{eq:sin} are the ones describing standard
Gaussian beams~\cite{siegman}: the plane wave phase $\eta$, the Gouy
phase $\psi_G \equiv \tan^{-1} \zeta$, the initial phase $\psi_0$ and
the phase advance due to the curvature of the wavefront: 
\be 
\psi_R =
\frac{kr^2}{2(z-z_0+\frac{z_R^2}{z-z_0})} 
\ee 
The overall phase $\psi$ is $\psi = \psi_0 +\eta -\psi_R
+\psi_G$. The fields have been computed analytically up to
$\epsilon^{11}$ in Ref.~\cite{salamin_e11}. In the study of the
optical cell described below we retain all terms up to $\epsilon^5$.

\begin{figure}
\centering\includegraphics[width=0.8\textwidth]{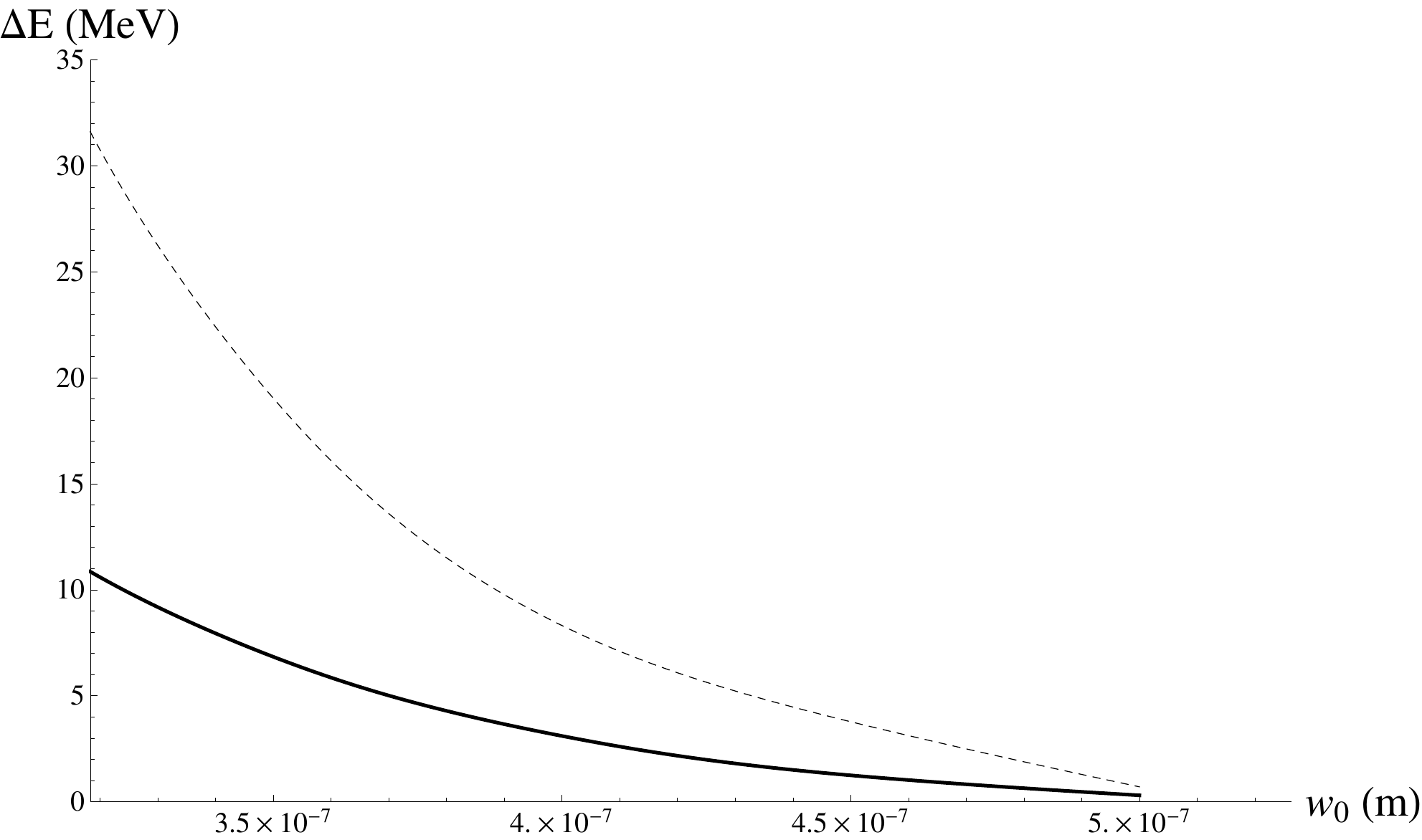} 
\caption{Net energy gain $\Delta E$ in MeV versus $w_0$ for a 100~MeV
  electron injected into the focus of a 1~TW (black continuous line) and 10~TW
  (dashed line) Gaussian beam ($\lambda = 1 \ \mu$m). The smallest
  value $w_0 = \lambda/\pi$ corresponds to the limit of the perturbative
  expansion of the fields ($\epsilon=1$). The largest value
  corresponds to $w_0= \lambda/2$.  }
\label{fig:energy_gain}
\end{figure}

A single pulse ($N=1$) optical cell can be operated even in the $\tau>T$
regime~\cite{esarey_1995,long_time} but in this case the particle must be injected
at an angle close to the $z$ direction. For $\tau>T$ acceleration is
mostly driven by $E_z$ and the ponderomotive force. Employing PW class
lasers~\cite{Mourou:2006zz} the energy gain can be very large although
the acceleration regime is highly non
linear~\cite{salamin_2002,Hsu:1997ib}. For $\tau<T$~\cite{lai_1980}
the particle of charge $q$ injected in the proximity of the $x$ axis
experiences only a portion of the $E_x$ period and net acceleration is
caused mostly by the $q  E_x$ linear term. The $B_y$
and $E_z$ fields contribute to the deflection angle in the $x-z$
plane. This can be demonstrated solving numerically the equation of
motion of the particle (electron, $q \equiv -e$) in the electric fields of
Eqs.~\ref{eq:ex}$\div$\ref{eq:bz}:
\be
\frac{d\mathbf{p}}{dt} = -e \left[ \mathbf{E} + c \boldsymbol{\beta}
\times \mathbf{B} \right] \ , 
\ \ \frac{dE}{dt} = -e c \boldsymbol{\beta} \cdot \mathbf{E}
\label{eq:pde_motion}
\ee
Numerical integration of Eq.~\ref{eq:pde_motion} is eased if the
motion is described in units of $T$ and the fields are expressed as
$\mathbf{\tilde{E}} \equiv (e/mc\omega) \mathbf{E}$ and
$\mathbf{\tilde{B}} \equiv (e/m\omega) \mathbf{B}$. The resulting
equation~\cite{salamin_2002} is:
\be 
\frac{d\boldsymbol{\beta}}{d\omega t} = \frac{1}{\gamma} \left[
  \boldsymbol{\beta}(\boldsymbol{\beta} \cdot \mathbf{\tilde{E}}) -
  (\mathbf{\tilde{E}} + \boldsymbol{\beta} \times \mathbf{\tilde{B}}) \right] .
\ee
In these formulas, $-e$ and $m$ are the charge and electron mass in SI
units, $\boldsymbol{\beta}$ is the electron velocity normalized to $c$
and $\gamma = (1-\beta^2)^{-1/2}$. Fig.~\ref{fig:energy_gain} shows
the net energy gain $\Delta E = (\gamma-\gamma_0)m c^2$ for a 100~MeV
electron ($\gamma_0 = 195.7$) injected along the $x$ axis as a
function of $w_0$. The initial phase $\psi_0$ is chosen in order to
have the electron at the center of the focus when $|E_x (\omega t)|$
reaches its maximum.  The laser pulse considered corresponds to 
$P_0$=1~TW and 10~TW power Gaussian beams focused to a beam spot
$w_0$. In Fig.~\ref{fig:energy_gain}, $w_0$ ranges from $w_0 =
\lambda/\pi$ to $\lambda/2$. The power crossing the disk of area $\pi
w_0^2$ at $z_0=0$ is $P_0(1-e^{-2}) \simeq 0.865 P_0$. The laser
intensity $I_0$ at $z=z_0$ and $r^2=0$ is proportional to the overall
beam power: $P_0 = \pi w_0^2 I_0/2$. The electric field is
\be
E_0 = \frac{1}{w_0} \sqrt{ \frac{4 P_0}{\pi \epsilon_0 c} } \ .
\ee 
For $w_0= \lambda/3$ and $P_0$=1~TW, $I_0$ is $5.7 \times 10^{20}$~W/cm$^2$ 
and $E_0 = 6.6 \times 10^{13}$~V/m. 

As expected, a net energy gain due to the electric field $E_x$ is
visible as long as the particle experiences just a fraction of the
oscillation period. In particular, no energy gain is observed when
$w_0> \lambda/2$ because the interaction time for an
ultra-relativistic particle $\tau \simeq 2 w_0/c$ is larger than an
optical cycle and hence, the electron experiences both an accelerating
and a decelerating electric field with nearly equal strength. For
$w_0< \lambda/2$ the energy gain $\Delta E$ grows linearly with $E_0$
and is proportional to $\sqrt{P_0}$. This scaling is in agreement with
results obtained under paraxial conditions~\cite{wong_2011} and in
non-paraxial treatments~\cite{marceau_2012,sepke_2006}. It is worth mentioning
that an increase of the laser wavelength at fixed $P_0$ does not
increase the energy gain per pulse for ultra-relativistic particles:
since $E_0 \sim w_0^{-1} \sim 1/\lambda$ and the interaction time
$\tau$ is $\sim 2w_0/c$, the momentum change experienced during the
pulse-electron interaction is $-e \int_0^\tau E_x dt \simeq -e \tau
E_0$ and it is thus independent of $\lambda$. Still, the exploitation of
mid-IR lasers ($\lambda \simeq 10$~$\mu$m), as $CO_2$
lasers~\cite{salamin_2011}, would be particularly rewarding in this
scheme because it would increase significantly the acceptance
of the cell. Progress toward sub-ps, 10 TW power $CO_2$ lasers are
very encouraging~\cite{pogorelsky_2010} but the capability to reach the
$w_0<\lambda/2$ condition in these lasers has not been demonstrated,
yet. 

\begin{figure}
\centering\includegraphics[width=0.8\textwidth]{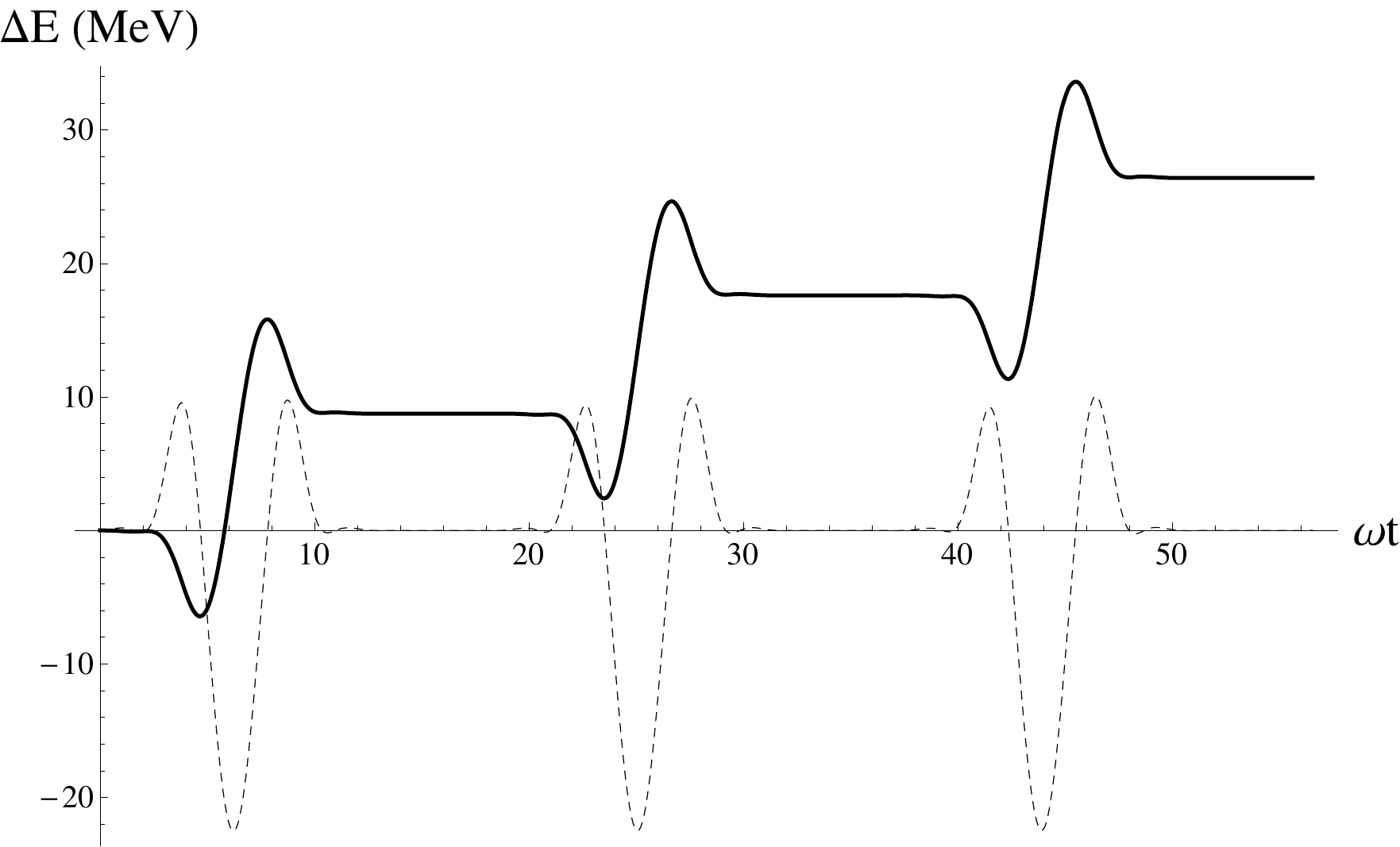} 
\caption{Black line: net energy gain $\Delta E$ in MeV versus $\omega t$ for a
  100~MeV electron injected into the focus of three 1~TW Gaussian
  beams ($\lambda = 1 \ \mu$m, $w_0= \lambda/3$). The accelerating
  field $\tilde{E_x}$ experienced by the particle at the time $\omega
  t$ is also shown (dashed line). }
\label{fig:enegain_three}
\end{figure}

The electron motion in the $z$-direction is determined by the
interplay between $B_y$ and $E_z$. In the proximity of the focus the
former boosts the electron toward the direction of motion of the laser
pulse, while for $z \rightarrow z_0$ and $r \rightarrow 0$, $E_z$ is
$\simeq 0$ and the effect of the longitudinal field $E_z$ is
marginal. Far from the focus, $B_y$ decreases steeply and the
contribution of $E_z$ becomes sizable. Fig.~\ref{fig:enegain_three}
shows the net energy gain $\Delta E$ for a 100~MeV electron injected
along $x$ at $z_0=0$ and crossing three pulses ($P_0 = 1$~TW, $w_0=
\lambda/3$) located at a distance $3\lambda$ in $x$ ($\Delta kx = 6
\pi$).  The accelerating field $\tilde{E_x}$ experienced along its
trajectory is also shown.

\begin{figure}
\centering\includegraphics[width=0.8\textwidth]{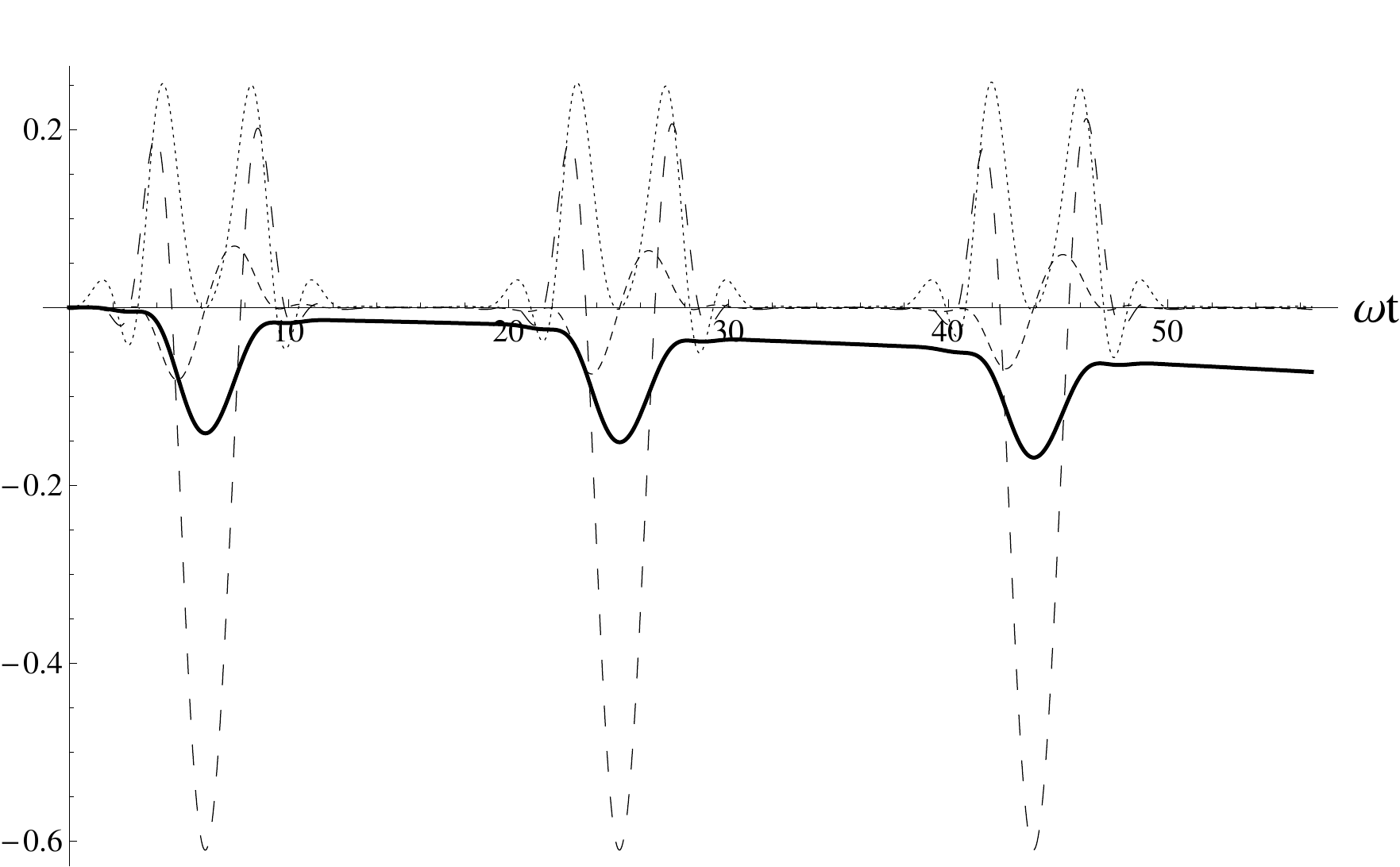} 
\caption{Position of the particle along $kz$ (black continuous line; $kz = 2\pi$
  corresponds to $z=\lambda$) and $\beta_z$ of the particle (short-dashed line)
  as a function of $\omega t$. Beam and laser parameters are the same
  as for Fig.~\ref{fig:enegain_three}.  The long-dashed and dotted lines
  show the values of $\tilde{B}_y/30$ and $\tilde{E}_z/30$
  experienced along the trajectory.  }
\label{fig:dkz_vs_time}
\end{figure}

The transverse motion of the particle is described in
Fig.~\ref{fig:dkz_vs_time}.  The continuous and short-dashed lines
indicate the position and $\beta_z$ of the particle along $kz$ ($kz =
2\pi$ corresponds to $z=\lambda$) as a function of $\omega t$. The
long-dashed and dotted lines show the values of the fields
$\tilde{B}_y$ and $\tilde{E}_z$ (divided by 30 to ease the reading of
the plot) experienced along the trajectory. For an ultra-relativistic
particle (100~MeV electron in Fig.~\ref{fig:dkz_vs_time}) starting at
$kx_0=-2 \pi$, the focus is reached at $\omega t = 2 \pi$. Far from
the focus, both $E_z$ and $B_y$ are positive and deflect the particle
toward negative values of $z$. Near the focus, $B_y$ is large and
negative and accelerate the electron along positive $z$. The net
result is that the two effects partially cancel when the electron
crosses the laser beam, and the overall deflection is much smaller
than one could envisage from the $\boldsymbol{\beta} \times
\mathbf{B}$ force only. This cancellation effect has been noted in the
90's and it is known to severely limits the energy transfer along $z$
in the $\tau > T$ regime~\cite{esarey_1995,sprangle}. If $\tau < T$,
however, the acceleration is due to $E_x$ and the cancellation above
helps to stabilize the particles in the transverse plane.  It is also worth
mentioning that the energy gain and the deflection of the electron is
correlated with the relative position and angle between the injected
electron and the laser. Besides acceleration, this feature might be
exploited for ultrafast diagnostics of sub-cycle laser pulses~\cite{referee}.

\begin{figure}
\centering\includegraphics[width=0.8\textwidth]{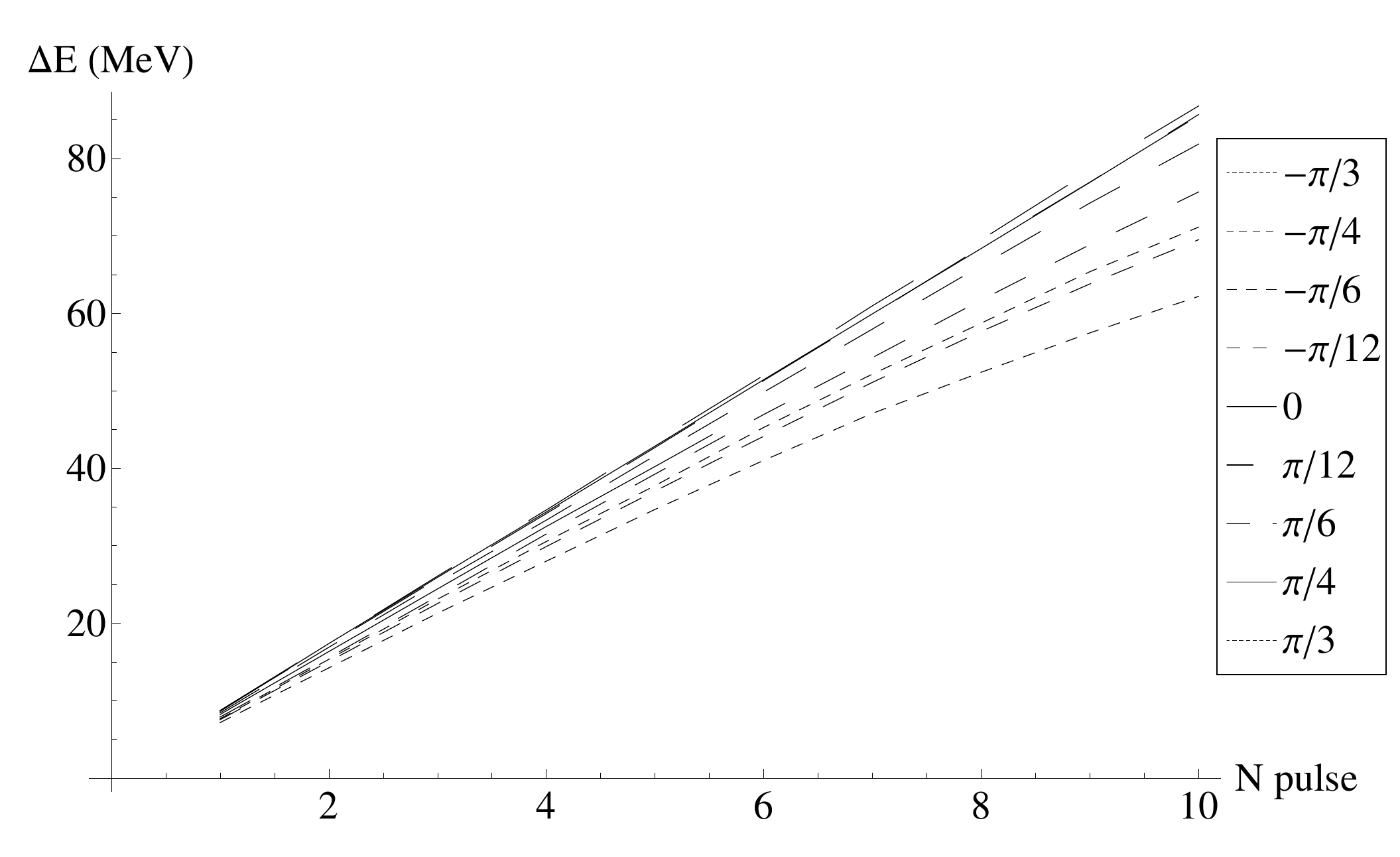} 
\caption{Energy gain $\Delta E$ (in MeV) after crossing $N$ laser
  pulses for 100~MeV electrons injected along $x$ at different
  positions in the $z$ axis.}
\label{fig:egain_vs_N}
\end{figure}

\begin{figure}
\centering\includegraphics[width=0.8\textwidth]{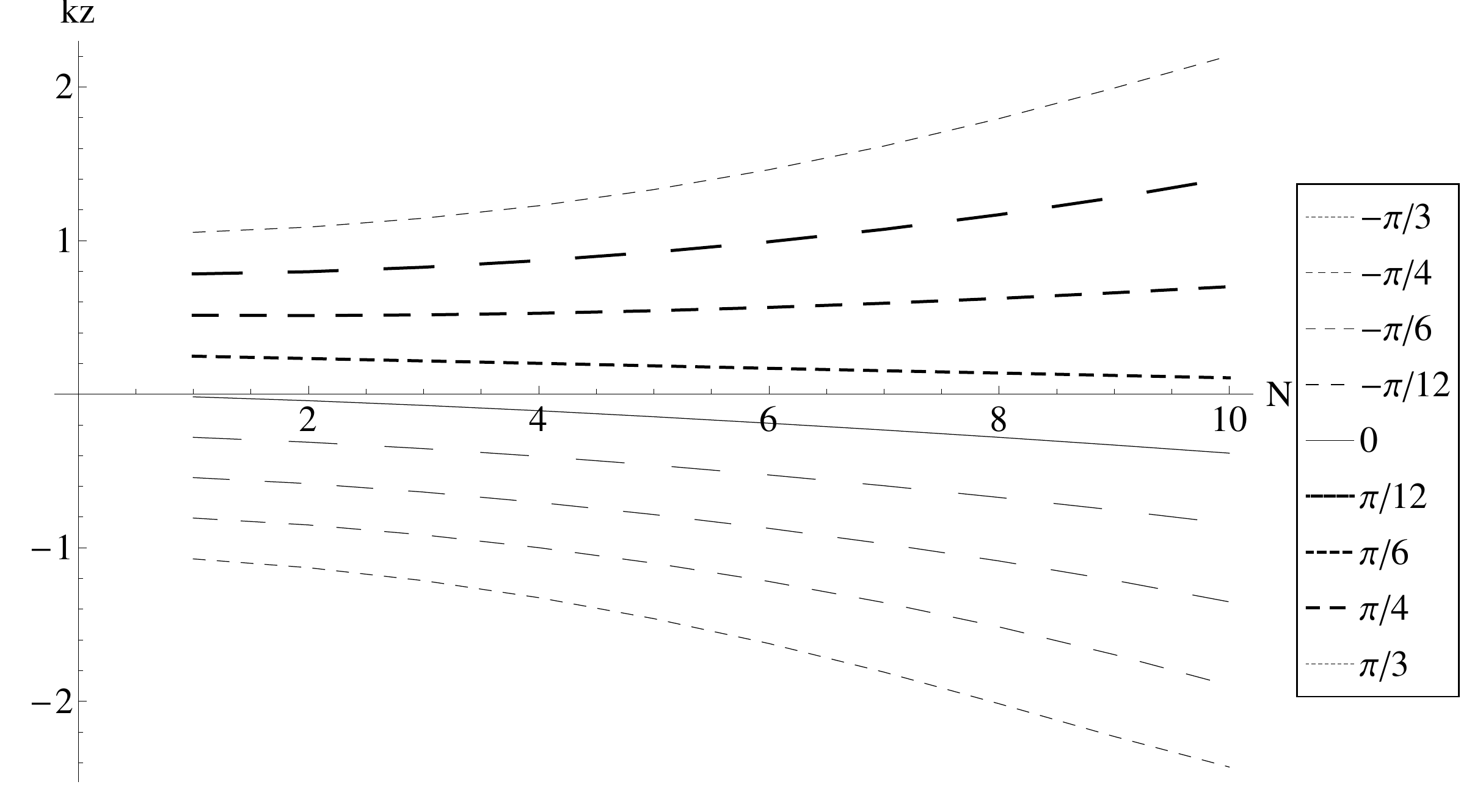} 
\caption{$kz$ position of the particles after crossing $N$
  laser pulses for 100~MeV electrons injected
  along $x$ at different positions in the $z$ axis.  }
\label{fig:kz_vs_N}
\end{figure}

In an optical cell of $N$ pulses, the final energy reached by the
particle and its trajectory in the transverse plane depends on the
initial position of the particle itself. If particles are
ultra-relativistic, they will always arrive in phase with the next
pulse, except for the slippage in $kx$ due to the fact that $\beta_z
\ne 0$ and the trajectory is no more rectilinear. The small distance
among the pulses and the cancellation effect of $B_y$ and $E_z$ in the
transverse plane make the phase slippage quite small and the transfer
function of the cell very regular. This is demonstrated
for $P_0=$1~TW, $w_0=\lambda/3$ pulses and 100~MeV electrons in
Figs.~\ref{fig:egain_vs_N} and
\ref{fig:kz_vs_N}. Fig.~\ref{fig:egain_vs_N} shows the energy gain as
a function of $N$ for particles injected at different $\Delta \equiv
kz(t=0)-kz_0$ (in the plot, $\Delta$ ranges between $-\pi/3$ and $\pi/3$).
Note that the maximum gain is observed for electrons that are injected
with a slight offset with respect to $z_0$ ($\Delta \simeq
\pi/12$). This is due to the residual bending in the transverse plane
experienced by particle injected directly at the focus ($\Delta=0$,
see Fig.~\ref{fig:dkz_vs_time}). It is also visible in
Fig.~\ref{fig:kz_vs_N}, where the $kz$ position of the particle after
$N$ electron-pulse interactions is shown. The energy spread at $N=10$
is 13\% and the maximum transverse distance is comparable to the
$\mathcal{O}(\lambda)$ size of the pulse in the $z$ direction.
Particles that are injected far from the focus will experience a
negligible acceleration and will flow unperturbed, while in-focus
electrons will experience acceleration (Fig.~\ref{fig:egain_vs_N}) and
defocusing (Fig.~\ref{fig:kz_vs_N}). The superposition of this
non-accelerated electron halo with in-focus electrons is a source of
non-laminar flow. To reduce the halo, it is therefore important that
the injected beam has a transverse size comparable with the laser beam
spot $w_0$. The finite contrast of the laser pulses, i.e. the presence
of pedestals along $z$ does not affect significantly the behavior of
the cell since pedestals do not overlap spatially with the accelerated
electron beam. Similarly, deviation from gaussianity of the pulse
envelope can affect the net energy gain per pulse $\Delta E$ through
the $-e \int_0^\tau E_x dt \simeq -e \tau E_0$ integral. In this case,
$\Delta E$ remains sizable if the $\tau < \lambda/c$ condition is
fulfilled.

The results of Figs.~\ref{fig:dkz_vs_time}-\ref{fig:kz_vs_N} are drawn
assuming ideal phase synchronization among the pulses and between
pulses and the incoming electrons. A systematic de-synchronization of
the pulses with respect to the injected electrons (``walk'') reduces
the overall gain and shifts the transfer function of Fig.6. For
instance, a 1 fs walk will reduce the energy gain by about
25\%. Random pulse-to-pulse phase variations that preserve the $\tau <
\lambda/c$ condition (e.g. a 1 fs ``jitter'' among pulses) average out
and will have no sizable effects in the operation of the cell.  In
fact, synchronization and carrier-envelope~\cite{moon_2009} phase
control of femtosecond pulses at the level that is requested for this
application are needed in other research fields too, and have been
achieved by attosecond pulse generators~\cite{chap3_chang}. Such a
level of stability control in laser pulses, however, has never been
demonstrated for the particular configuration of the optical cell,
where sub-fs synchronization and the $w_0<\lambda/2$ condition have to
be fulfilled for all pulses confined within the cell. Similarly,
beam-laser synchronization techniques developed for
free-electron-lasers (FEL) and external-injection laser-wakefield
acceleration~\cite{sparc_lab} are aimed to a precision of
$\mathcal{O}(10)$~fs. Those beams can be employed for a
proof-of-principle of this acceleration scheme but the level of
synchronization is still significantly larger than the half-period
scale (1.7~fs for $\lambda=1$~$\mu$m) needed for a fully efficient
exploitation of the optical cell. For the beam parameters of
\cite{sparc_lab} the ratio ($R$) between the energy transferred to the
particles through the cell and the total laser pulse energy stored in
the cell is $R<0.1$. In this regime the effects of laser depletion,
distortion and space charge can be neglected, while the study of the
high beam loading regime ($R \simeq 1$) requires a full multiparticle
simulation. Similarly, a multiparticle simulation of the whole system of
optical cells and refocusing units would be needed to study the
occurrence of beam break-up instabilities. Still, the small size of
the cell and the absence of plasma sheaths make this scheme less
sensitive to such class of instabilities and to the resonant growth of
head-tail oscillations inside the optical cell.
 
Fig.~\ref{fig:kz_vs_N} shows that the cell defocuses the electrons in
the $x-z$ plane.  It is, thus, inconvenient to build optical cells
with $N \gg 10$ before proceeding to particle refocusing because
peripheral particles will not experience large values of $E_x$ at the
end of the cell. Note also that conventional refocusing units have
lengths of ${\cal O}$(1 m). The minimum distance between optical cells is thus
much larger than the longitudinal size of the cell.
The overall energy gain of an optical cell is
independent of the initial energy of the electron as far as the
particle is ultra-relativistic but, as for standard lenses, the cell
causes chromatic aberration. For a 1~GeV initial energy electron the
$N_0=10$ cell accelerates particles up to $\Delta E\simeq 80$~MeV but,
in this case, the spread in the transverse plane is $\sim$10 times
smaller because the magnetic rigidity is proportional to the particle
momentum. At larger momenta and ultra-relativistic particles ($E
\simeq pc$), it is thus appropriate to employ cells with $N \simeq N_0
\cdot p(\mathrm{MeV})/100$.

\begin{figure}
\centering\includegraphics[width=0.8\textwidth]{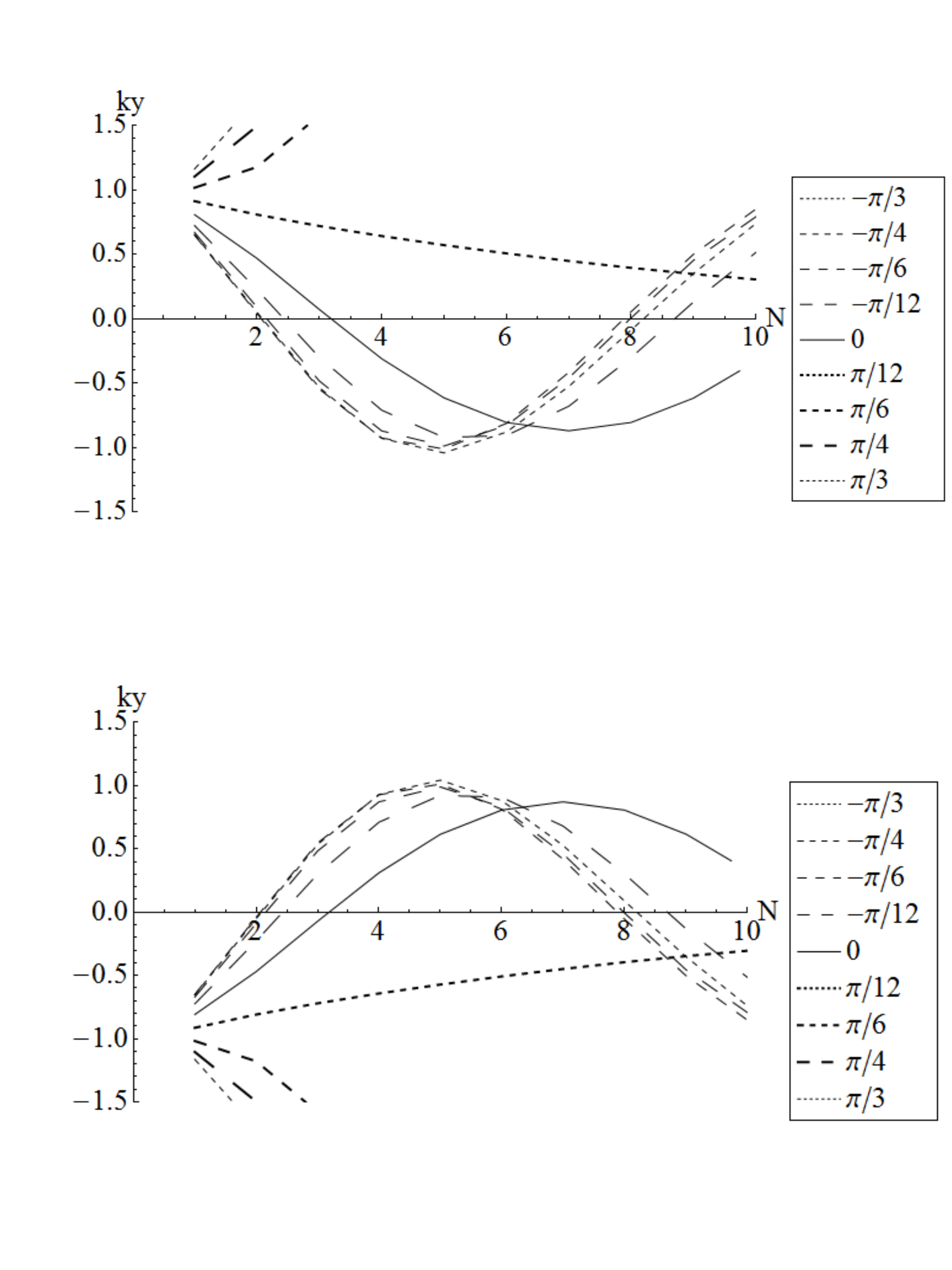} 
\caption{$ky$ position of the particles after crossing $N$ laser
  pulses for 100~MeV electrons injected along $x$ at different
  positions in the $z$ axis. The electrons are injected with an offset
  in $y$ equal to $\pi/3$ (top plot) and to $-\pi/3$ (bottom plot).  }
\label{fig:ydispl}
\end{figure}

Particles that are injected close to the $y=0$ axis at any position in
$z$ will remain in the $x-z$ plane.  On the other hand, if they are
significantly displaced in $y$ ($\Delta ky = \pm \pi/3$ in
Fig.~\ref{fig:ydispl}) they will experience the fields near pulse
boundaries. In general, these particles will oscillate around the $y=0$ axis,
but electrons that are also displaced in the $kz>0$ direction will be
driven outside the cell acceptance. This is shown in
Fig.~\ref{fig:ydispl} where the $ky$ position of the particles are
traced as a function of the number of crossed laser pulses.
Fig.~\ref{fig:ydispl} refers to 100~MeV electrons injected along $x$
at different positions in the $z$ axis with an offset $\Delta ky$ in
the $y$ direction equal to $\pi/3$ (top plot) and to $-\pi/3$ (bottom
plot).


In conclusion, even the simplest configuration of the optical cell can
be operated with ultra-relativistic particles in order to increase the
energy to an amount proportional to $N$. The energy gain is sizable if
the spot size is smaller than $\lambda/2$ and scales as $P_0^{1/2}$.
The optical cell can be operated with $N \simeq 10$ for 100~MeV
electrons and the optimum number of pulses increases linearly with the
initial momentum of the particles. 
The acceptance of the
optical cell, i.e. the region in the transverse plane where particles
can experience acceleration, is of the order of $w_0$. From a
practical point of view, it represents the most notable limitation
of this scheme: stable and efficient acceleration of particles
requires injected beams of $\mu$m size for near-infrared lasers and
beam-laser synchronization at the sub-fs level. As mentioned above,
beam manipulation techniques developed for FEL allow for a
proof-of-principle of the optical cell concept but high transport and
acceleration efficiency through optical cells remains a major
technological challenge.

\section*{Acknowledgments}

The author gratefully acknowledges comments and suggestions from
S.V.~Bulanov, R.~Felici, K.T.~McDonald and F.~Pegoraro.


\end{document}